\documentclass[aps,twocolumn,showpacs,amsmath,amssymb]{revtex4-1}
\usepackage{graphicx}

\begin{document}
\title{Mechanism of destruction of transport barriers in
geophysical jets with Rossby waves}

\author{M.Yu. Uleysky, M.V. Budyansky, and S.V. Prants}
\affiliation{Pacific Oceanological Institute \\
of the Russian Academy of Sciences, 43 Baltiiskaya st., 690041 Vladivostok,
Russia}
\begin{abstract}
The mechanism of destruction of a central transport barrier
in a dynamical model of a geophysical zonal jet current in the ocean
or the atmosphere with two propagating Rossby waves is studied.
We develop a method for computing a central invariant
curve which is an indicator of existence of the barrier. 
Breakdown of this curve under a variation
of the Rossby wave amplitudes and onset of chaotic cross-jet
transport happen due to specific resonances
producing stochastic layers in the central jet. The main result is that 
there are resonances breaking 
the transport barrier at unexpectedly small values of the amplitudes that 
may have serious impact on mixing and transport in the ocean and the 
atmosphere. The effect can be found in laboratory experiments with azimuthal 
jets and Rossby waves in rotating tanks under specific values of the 
wave numbers that are predicted in the theory.
\end{abstract}
\maketitle

Transport and mixing of water (air) masses and their characteristics play
a crucial role in the ocean and atmosphere dynamics. In the Lagrangian
approach a particle with the position $\vec r$ is advected by an Eulerian
velocity field $\vec v(\vec r, t)$
\begin{equation}
\frac{d \vec r}{d t}=\vec v(\vec r, t).
\label{1}
\end{equation}
It is known that a simple deterministic velocity field may produce
practically unpredictable particle trajectories, the phenomenon
known as chaotic advection \cite{Aref, Ottino, PK06}.

We study theoretically and numerically horizontal cross-jet transport
in geophysical zonal flows. To list a few we mention the Gulf Stream
in the Atlantic, the Kuroshio in the Pacific, and the polar night
Antarctic jet in the atmosphere, which are the jet currents separating 
water (air)
masses with different physical properties. Transport of particles
across a geophysical jet is of crucial importance and may cause, for 
example, depletion of ozone
in the atmosphere and heating and freshing of waters in the ocean. The velocity
fields of real flows are not, of course, regular, but if the Eulerian
correlation time is large as compared to the Lagrangian one, the problem may
be treated in the framework of chaotic advection concept.

The equations of motion of a passive particle with coordinates $x$ and $y$ 
advected by a two-dimensional
incompressible flow with a stream function $\Psi$ 
are known to have a Hamiltonian form
\cite{Aref}
\begin{equation}
\frac{d x}{d t}=u(x,y,t)=-\frac{\partial \Psi}{\partial y},\quad
\frac{d y}{d t}=v(x,y,t)=\frac{\partial \Psi}{\partial x},
\label{2}
\end{equation}
with the phase space being the position space for advected particles.
Chaotic mixing and transport in jet flows have been extensively
studied with kinematic models, where the velocity field is a given
function of $x$, $y$ and $t$ imitating real flows
(see \cite{PK06,S92,RWig06,UBP07} and references therein), and with
dynamical models conserving the potential vorticity
(see \cite{PK06,P91,DM93,Rypina} and references therein).
The problem has been studied as well
in laboratory where azimuthal jets with Rossby waves have been 
produced in rotating tanks \cite{SMS89,SHS93}.
It has been found both numerically and experimentally  that fluid
is effectively mixed along the jet, but in common opinion a large
gradient of the potential vorticity in the central part prevents
transport across the jet. A technique, based on computing the finite-scale
Lyapunov exponent, 
has been found useful in Ref.~\cite{BLR01} to detect the presence of
cross-jet barriers in kinematic models. A comparison of properties
of cross-jet transport in kinematic and dynamical models of atmospheric 
zonal jets has been done recently in Ref.~\cite{P.H.Haynes}. 
Up to now, the transport barrier has been shown 
numerically \cite{Rypina,RWig06} to be broken only with so large values 
of the wave amplitudes that are beyond of the validity of linear 
models and can be hardly observed in real flows.

The aim of the paper is to prove that cross-jet transport under
appropriate conditions is possible at comparatively small values of the wave
amplitudes and, therefore, may occur in geophysical jets. We develop 
a general method to detect a core of the transport
barrier and find a mechanism of its destruction using the dynamical
model of a zonal jet flow with two propagating Rossby waves. The
method comprises the identification of a central invariant curve (CIC),
which is an indicator of existence of the barrier, finding certain resonant 
conditions for its destruction at
given values of the wave numbers, and detection of cross-jet transport.

Motion of two-dimensional incompressible fluid in the rotating frame is
governed by the equation for conserving potential vorticity 
$(\partial/\partial t+\vec v\cdot \vec \nabla)\Pi=0$. In the 
quasigeostrophic approximation \cite{P87}, one gets $\Pi=\nabla^2\Psi+\beta y$,
where $\beta$ is the Coriolis parameter. The $x$ axis is chosen
along the zonal flow, from the west to the east and $y$~--- along the gradient
from the south to the north. Barotropic perturbations of zonal flows
produce Rossby waves which have an essential impact on transport
and mixing in the ocean and the atmosphere \cite{P87}.
The stream function is sought in the form
\begin{equation}
\Psi=\Psi_0+\Psi_\text{int}=\Psi_0(y)+\sum_j\Phi_j(y) e^{ik_j(x-c_jt)},
\label{3}
\end{equation}
where $\Psi_0$ describes a zonal flow and $\Psi_\text{int}$ is its perturbation
which is supposed to be a superposition of zonal running Rossby waves. After
substituting (\ref{3}) in the equation for the potential vorticity and
a linearization, one gets the Rayleigh-Kuo equation \cite{Kuo}
\begin{equation}
(u_0-c_j)\Bigl(\frac{d^2 \Phi_j}{d y^2}-k_j^2\Phi_j\Bigr)+
\Bigl(\beta-\frac{d^2 u_0}{d y^2} \Bigr)\Phi_j=0,
\label{4}
\end{equation}
where the zonal velocity $u_0=-d \Psi_0/d y$ has a single extremum
at $y=0$. If one takes the following zonal-velocity profile 
(the Bickley jet \cite{DM93}): 
\begin{equation}
u_0(y)=U_0\operatorname{sech}^2{\frac{y}{D}},
\label{5}
\end{equation}
then Eq.~(\ref{4}) admits two neutrally stable solutions
\begin{equation}
\Phi_j(y)=A_j U_0 D\operatorname{sech}^2{\frac{y}{D}},\quad j=1,2,
\label{6}
\end{equation}
where $U_0$ is the maximal velocity in the flow, $D$ is a measure of its width, and
$A_j$ are the wave amplitudes. It is easy to check that
(\ref{5}) and (\ref{6}) are compatible with (\ref{4}) if there is the
following condition for the phase velocities:
\begin{equation}
c_{1,2}=\frac{U_0}{3}(1\pm\alpha),\quad
\alpha \equiv \sqrt{1-\beta^{\ast}},\quad
\beta^\ast\equiv\frac{3 D^2 \beta}{2 U_0},
\label{7}
\end{equation}
which are connected with the wave numbers by the dispersion
relation $c_{1,2}=U_0 D^2 k_{1,2}^2/6$. Two neutrally stable
Rossby waves exist if $\beta D^2/U_0<2/3$.

Thus, the stream function of the zonal flow  with two
Rossby waves, satisfying the conservation
of the potential vorticity, has the form 
\begin{multline}
\Psi (x,y,t)=-U_0 D\Bigl(\tanh{\frac{y}{D}}-\operatorname{sech}^2
{\frac{y}{D}}\times\\
\times\Bigl[A_1 \cos{k_1(x-c_1 t)}+
A_2 \cos{k_2(x-c_2 t)}\Bigr]\Bigr).
\label{10}
\end{multline}
One of the task of this paper is to present results in the form
allowing a comparison with laboratory experiments \cite{SMS89,SHS93}
in which an azimuthal jet at the radius $R$ with Rossby waves with the 
wave numbers $n_1$ and $n_2$ has been produced:
\begin{equation}
k_{1,2}=\frac{n_{1,2}}{R},\quad c_{1,2}=\frac{U_0 D^2}{6 R^2} n_{1,2}^2.
\label{11}
\end{equation}
Let it be $n_1>n_2$, and the wave with $n_1$ is called the first one. Let
the wave numbers be represented as $n_1=m N_1$ and $n_2=m N_2$,
where $m\ne 1$ is the greatest common divisor and $N_1/N_2$ is an
irreducible fraction. Introducing new coordinates $x'$, $y'$, and $t'$
\begin{equation}
x=\frac{(x'+C_2t')R}{m},\quad y=Dy',\quad t=\frac{R}{mU_0}t',
\label{13}
\end{equation}
we rewrite the stream function (\ref{10}) in the frame moving with the phase
velocity of the first wave 
%
%\begin{equation}
\begin{multline}
\Psi'(x',y',t')=-\tanh{y'}+A_1\operatorname{sech}^2{y'}\cos (N_1 x') +\\
+A_2\operatorname{sech}^2{y'}\cos (N_2 x'+\omega_2 t') + C_2y',
\label{14}
\end{multline}
%\end{equation}
%
where
\begin{equation}
\omega_2\equiv\frac{2 N_2 (N_1^2-N_2^2)}{3(N_1^2+N_2^2)},\quad
C_2\equiv \frac{2N_1^2}{3(N_1^2+N_2^2)}.
\label{15}
\end{equation}
Thus, we get the stream function (\ref{14}) with the control parameters
$N_1$ and $N_2$ which are specified by the four experimental
parameters: $U_0$, $\beta$, $D$, and $R$. One can now study cross-jet
transport with any combination of the wave numbers $n_1$ and $n_2$
that can be realized in a laboratory experiment by adjusting
the radius $R$, the jet width $D$, the maximal velocity $U_0$,
and the Coriolis-like parameter $\beta$ \cite{SMS89,SHS93}. 

\begin{figure}[!htb]
\includegraphics[width=0.48\textwidth]{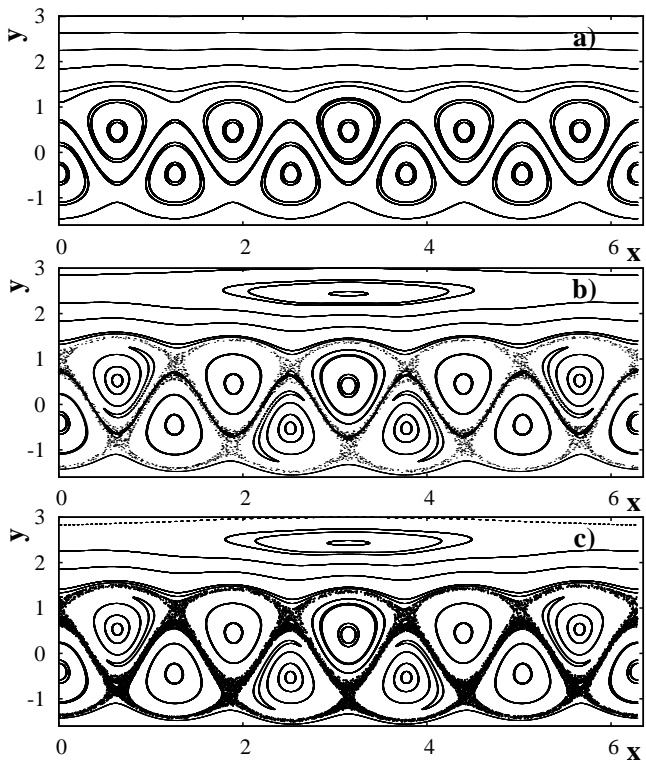}
\caption{(a) $A_2=0$. Phase portrait of the steady jet flow with
$N_1=5$ and $N_2=1$ in the moving frame. (b) $A_2=0.09$. CIC (the bold curve) 
is a barrier to transport across the jet. (c) $A_2=0.095$. Destruction of CIC
and onset of cross-jet transport.}
\label{fig1}
\end{figure}

The advection equations (\ref{2}) with the stream function
(\ref{14}) have the form
\begin{equation}
\begin{gathered}
\begin{aligned}
\frac{d x}{d t}=&-C_2+\operatorname{sech}^2{y}[1+2A_1\tanh{y}\cos{(N_1 x)}+\\
&+2A_2\tanh{y}\cos{(N_2 x+\omega_2 t)}],
\end{aligned}\\
\frac{d y}{d t}=-\operatorname{sech}^2{y}[A_1 N_1\sin{(N_1 x)}+
A_2 N_2\sin{(N_2 x+\omega_2 t)}],
\end{gathered}
\label{16}
\end{equation}
where we omitted the primes over $x$, $y$, and $t$. 
If the amplitude of the second wave is zero, $A_2=0$, then the set
(\ref{16}) is integrable. The phase portrait of the steady flow
with $A_1=0.2416$, $N_1=5$ and $N_2=1$ is shown in Fig.~\ref{fig1}a in the frame
moving with the phase velocity of the first wave. The
eastward jet is situated between two chains with five vortices. 
The southern and northern peripheral currents are westward in the moving frame. 
In a steady flow all the particles follow streamlines.  At $A_2>0$,
chaos may arise in a typical way: a stochastic layer appears at the place
of the broken separatrices (Fig.~\ref{fig1}b and c).

At odd values of $N_1$ and $N_2$, Eqs.~(\ref{16}) have the two
symmetries
\begin{equation}
\hat S:\left\{
\begin{aligned}
\tilde x &=\pi+x,\\
\tilde y &=-y,
\end{aligned}\right.
\quad
\hat I_0:\left\{
\begin{aligned}
\tilde x &=-x,\\
\tilde y &=y,
\end{aligned}\right.
\label{**}
\end{equation}
which are involutions, i.~e., $\hat S^2=1$ and $\hat I_0^2=1$.
Solving the equation $\hat I_0(x_j, y_j)=\hat S(x_j, y_j)$, $j=1,2$,
one gets indicator points \cite{Aizawa}: ($x_1=\pi/2$, $y_1=0$)
and ($x_2=3\pi/2$, $y_2=0$).
Iterating them, we construct a CIC \cite{BUP09} in the central part of the
jet which is the last transport barrier in the sense that the CIC breaks
down and is replaced by a stochastic layer with variation of the wave
amplitudes. We illustrate this in Fig.~\ref{fig1}. At $A_2=0.09$, the CIC together with neighboring
invariant curves forms a narrow transport barrier (Fig.~\ref{fig1}b).
We define a CIC as a curve which is invariant under the operator $\hat S$ and the 
evolution operator over the period $2\pi/\omega_2$. The CIC separates the 
northern and southern parts of the flow.  
At $A_2=0.095$, the CIC breaks down, and cross-jet transport becomes
possible (Fig.~\ref{fig1}c).

It is reasonable to suppose that destruction of CIC is caused by
a ballistic resonance between the maximal frequency of the particle motion
in the central jet and the perturbation frequency $\omega_2$. The first one is
estimated from Eq.~(\ref{16}) to be $f_1\simeq -C_2+1$, and the second
one is given by (\ref{15}). So, the approximate condition of the ballistic
resonance is 
\begin{equation}
\frac{f_1}{\omega_2}=\frac{N_1^2+3N_2^2}{2N_2(N_1^2-N_2^2)}.
\label{19}
\end{equation}
At small amplitudes, this ratio gives an approximate estimate for the CIC 
winding number $w$ \cite{BUP09}. Equating the right-hand side of Eq.~(\ref{19}) 
to a rational number, one finds those values of the wave numbers $N_1$ and $N_2$ 
for which the CIC is strongly influenced  by the corresponding resonance, and, 
therefore, cross-jet transport becomes possible.
\begin{figure}[!htb]
\includegraphics[width=0.40\textwidth]{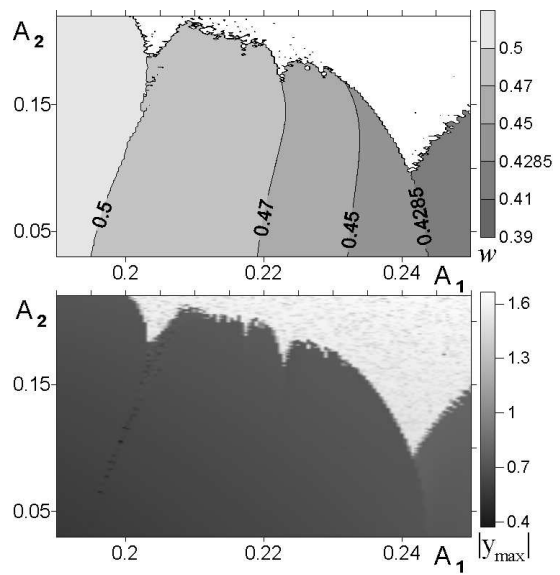}
\caption{Diagrams of (a) the winding number $w$ and (b) the maximal deviation 
of iterations of the indicator point along the y-axis,  $|y_{\rm max}|$, 
in the space of the Rossby wave amplitudes $A_1$ and $A_2$.
White zone: regime with cross-jet transport.}
\label{fig2}
\end{figure}

In order to reveal a scenario for CIC destruction we plot in Fig.~\ref{fig2}
the dependencies of $w$ and the maximal deviation 
of iterations of the indicator point along the y-axis,  $|y_{\rm max}|$, on $A_1$ and $A_2$
for the pair ($N_1=5$, $N_2=1$). The bifurcation curves
with the winding numbers corresponding to certain resonances
are shown in Fig.~\ref{fig2}a. The even $2:1$ ($w=0.5$) and odd $7:3$
($w=0.4285$) ones produce two deep and wide spikes in the plots.
White color codes those values of the amplitudes $A_{1,2}$ at
which a CIC is broken. Comparing Figs.~\ref{fig2}a and b, we see that 
the zone with broken CIC in Fig.~\ref{fig2}a 
correspond to the values $|y_{\rm max}| \simeq 1.5$ in Fig.~\ref{fig2}b,  
i.e., iterations of points, situated initially in the jet core, 
cover the region of the size of order $\simeq 3$ jet's half-width. 
It means breakdown of central transport barrier at those values of 
$A_{1,2}$ at which the CIC is broken.
Figure~2 demonstrates clearly that destruction of the transport barrier 
may happen at comparatively small values of the wave amplitudes  
$(A_{1,2} < 1)$. Our model is essentially a linear one, and the 
Rayleigh-Kuo equation is valid to first order in the wave amplitudes. 

\begin{figure}[!htb]
\includegraphics[width=0.43\textwidth]{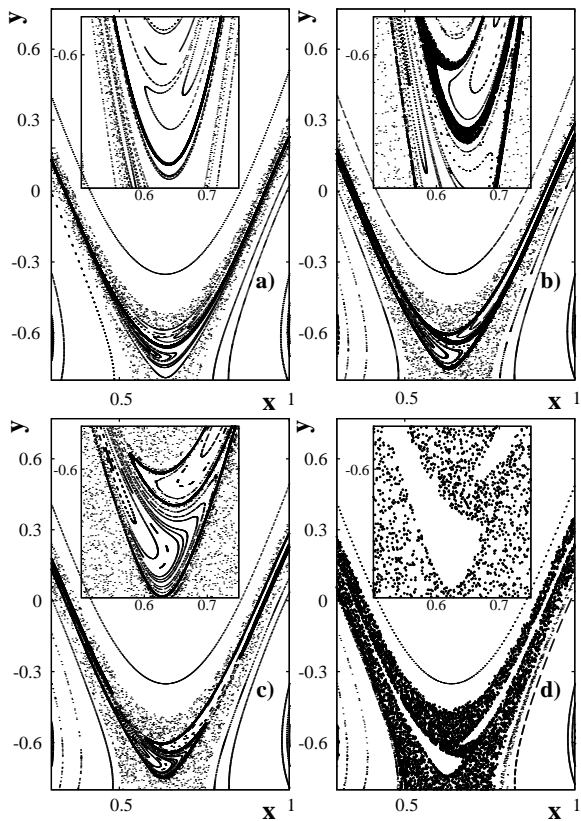}
\caption{Mechanism of CIC destruction and onset of cross-jet transport. 
(a) $A_2=0.087$. Smooth CIC and neighboring invariant curves form a transport
barrier. (b) $A_2=0.088$. A narrow stochastic layer (shadowed strip) appears at the 
place of the broken CIC. (c) $A_2=0.09$. CIC appears again as a meandering 
curve. (d) $A_2=0.1$. Breakdown of CIC and onset of cross-jet transport. 
Insets show magnification of
the phase space region nearby the resonance $7:3$.}
\label{fig3}
\end{figure}

To illustrate the mechanism of destruction of CIC we study the topology
of the phase space near the islands of the resonance $7:3$
(see the spike with $w=0.4285\dots$ in Fig.~\ref{fig2}).
Let us fix $A_1=0.2418$ and gradually increase $A_2$ away from zero.
In the range $0<A_2<0.088$ the smooth CIC and neighboring invariant
curves form a transport barrier (Fig.~\ref{fig3}a). At $A_2 \simeq 0.088$,
invariant manifolds of hyperbolic orbits of the resonance $7:3$
cross each other, the CIC breaks down, and there appears at its
place a narrow stochastic layer locked between remained invariant 
curves (Fig.~\ref{fig3}b). When $A_2$ increases further islands of
the resonance $7:3$ diverge, and a meandering CIC appears again 
between them (see Fig.~\ref{fig3}c at $A_2=0.09$). At
$A_2>0.095$, CIC and surrounding invariant curves are destroyed,
and cross-jet transport becomes possible in a wide range of the
$y$ coordinate (Fig.~\ref{fig3}). Thus, existence of a CIC is a sufficient  
but not necessary condition for existence of a transport barrier.
Animation of metamorphoses of topology
of the transport barrier and its destruction at a fixed value of one
of the wave amplitudes and variation of the other one can be found
in Ref.~\cite{animation}.

In conclusion we discuss briefly a possibility for checking main
results of our work in laboratory experiments on chaotic advection
in rotating fluid \cite{SMS89,SHS93} imitating nonlinear geostrophical 
geophysical flows
in the ocean and the atmosphere. An azimuthal jet with Rossby waves
was produced by the action of the Coriolis force on radially
pumped fluid in a rotating tank with a slope imitating the $\beta$-effect on 
the rotating Earth. The measured velocity field was
rather well approximated by the model stream function
(\ref{10}) \cite{SHS93}. Rapid mixing on either side
of the jet was observed for a  quasiperiodic flow, but no significant
transport was observed across the jet. In our opinion the reason
is that the experiments have been carried out under conditions 
that are far away from the resonances which are capable 
of destroying the central transport barrier at the values of the Rossby wave
numbers realized in the experiment. The results obtained in this paper 
allow to specify those values of the control parameters of the flow, 
the Rossby wave numbers, for which there exist specific resonances 
capable of destroying transport barriers at comparatively small 
values of the wave amplitudes. Our recommendation to observe
cross-jet transport in laboratory is to produce an azimuthal jet and 
Rossby waves with odd wave numbers whose values differ significantly from
each other, say ($N_1=5, N_2=1$) or ($N_1=7, N_2=3$).

The work was supported partially by the Program
``Fundamental Problems of  Nonlinear Dynamics'' of the Russian
Academy of Sciences and by the Russian Foundation  
for Basic Research (project no. 09-05-98520).

\end{document}